\documentclass[envcountsect]{article}

\usepackage[utf8]{inputenc}
\usepackage[T1]{fontenc}

\usepackage{amsmath,amsfonts,amssymb, amsthm}
\usepackage{url}
\usepackage{hyperref}
\usepackage{authblk}

 \newcommand\R{\mathbb{R}}

 \newcommand\Z{\mathbb{Z}}

 \newcommand{\ket}[1]{\left\vert#1\right\rangle}

 \newcommand\floor[1]{\left\lfloor #1\right\rfloor}

\begin{document}

\title{Offloading Quantum Computation by Superposition Masking}

\author[1]{Samuel Jaques\thanks{Supported by the University of Oxford Clarendon Fund}}
\affil[1]{Department of Materials, University of Oxford, UK samuel.jaques@materials.ox.ac.uk}
\author[2]{Craig Gidney}
\affil[2]{Google Inc., Santa Barbara, California 93117, USA}
\date{}

\maketitle

\begin{abstract}
    Error correction will add so much overhead to large quantum computations that we suspect the most efficient algorithms will use a classical co-processor to do as much work as possible. We present a method to offload portions of a quantum computation to a classical computer by producing a superposition of masks which hide a quantum input. With the masks, we can measure the result without altering the original input and then perform classical computations on the measured output. If the task has enough structure, the classical computations will be equivalent to a quantum computation performed in superposition. We apply this technique to modular inversion, root-finding, division with remainder, sparse matrix inversion, and inverting generic group homomorphisms, achieving at least a constant-factor improvement in quantum operations for each. Unfortunately, it is difficult to uncompute or invert this technique because of the measurement, and thus we know of no useful algorithm which benefits from superposition masking.
\end{abstract}

\section{Introduction}
Quantum algorithms have asymptotic advantanges over classical algorithms on certain problems~\cite{grover_1996,shor_1994}.
However, these asymptotic results ignore constant factors and quantum computers are expected to have non-negligible constant factor penalties relative to classical computers due to the overhead of error correction~\cite{FMMC2012surface_code,gheorghiu2019benchmarking,gidney2019factor,suchara_et_al_2013}.
In fact, the expected constant factor difference is large enough that it is worth spending some time emphasizing.

Classically, performing a logic gate is too cheap to count.
A single CPU, with its billions of transistors and clock speeds in the gigahertz, performs quintillions of logic gates per second.
Quantumly, individual logic gates are too expensive to ignore.
For example, suppose we have a large scale quantum computer based on applying the surface code to superconducting qubits (with a physical gate error rate around one in a thousand, a surface code cycle time around a microsecond, and a desired logical gate error rate below one in a billion).
In this context, a single quantum CNOT gate would involve thousands of physical qubits and take tens of microseconds~\cite{horsman_surface_2012}.
Real time error correction of the physical measurements performed as part of the CNOT could saturate multiple CPUs~\cite{FWH2012surface_code,TMNCQM2017QECC_hardware}.
This suggests that the constant factor of running a gate quantumly, instead of classically, is at least one billion (in terms of area time).

The huge constant factor penalty of quantum computation over classical computation creates an interesting dynamic in the design of efficient quantum computations. Anything that can be offloaded to a classical computer is effectively free.
As an example, a classical computer can compute a table of data and the quantum computer can do lookups in the table to accelerate operations such as big integer multiplication \cite{gidney2019windowed,vanmeter2005exp}. These techniques led to the cheapest methods for Shor's algorithm \cite{gidney2019factor,haner_improved_2020}.

Another interesting dynamic which emerges from the design of efficient quantum computations is an asymmetry between computation and uncomputation.
For example, computing the AND of two qubits requires non-stabilizer operations such as T gates or Toffoli gates but the AND can be uncomputed using only stabilizer operations~\cite{jones_2013}.
In the surface code, non-stabilizer operations requires costly operations such as magic state distillation \cite{bravyi_kitaev_2004,SciAdv:Brown2020,FMMC2012surface_code}, and as a result the area time of an AND gate computation is an order of magnitude or two larger than for its uncomputation.
The asymmetry in the cost ultimately comes from the irreversibility of measurement.
The uncomputation of the AND gate uses measurement in a crucial way, but because measurement has no inverse this trick can't be inverted for use during the initial computation of the AND gate.

Interestingly, in all prior examples of compute/uncompute cost asymmetry that we are aware of~\cite{Berry2019qubitizationof, gidney2019windowed, jones_2013}, it is the uncomputation that benefits from the ability to perform measurement.
Our results in this paper are the first example we know of where the ability to measure favors the computation.

Section~\ref{sec:examples} gives an overview of the technique, which we call superposition masking. Instead of a rigorous generalization, we give several example applications. We show both constant and non-constant improvements in quantum costs, at the expense of more classical computation. Given the large constant factor difference between quantum and classical computation, we expect that our method improves the real cost of these computations.

In Section~\ref{sec:discussion} we discuss limitations and complexity. While some of our examples provide asymptotic improvements in quantum gate cost, to the best of our knowledge, none of these techniques help with a \emph{specific} quantum algorithm. Our hope is that the examples convey the sense of the idea, and that there will be some application in the future that can use this technique.

\section{Superposition Masking}\label{sec:examples}
\subsection{Overview}
Most quantum algorithms involve some fully-quantum step such as a controlled phase shift or a quantum Fourier transform, punctuated by a quantum implementation of a classical task. The fully-quantum steps are often a negligible fraction of the entire cost. Thus, we would like to perform as much of the classical task as possible on a classical co-processor. 

For the classical processor to process the quantum data, it would need to measure it, which would destroy the state. To preserve the data through measurement, we introduce superposition masking. In each application, we follow the same steps:
\begin{enumerate}
    \item 
    Create a superposition of masks in a new register. 
    \item
    Combine the masks with the input in some way and measure the result.
    \item
    Perform an expensive classical computation on the measurement result.
    \item
    Write the output of this computation into a quantum register, and clean up the mask.
\end{enumerate}

For this to work properly, we need to ensure that when we combine the masks with the input, the result is completely independent of the input, so that the state is preserved when we measure. Then the measurement just entangles the masks with the inputs. 

The classical computation can only operate on one input, so we can only use this method for functions that are sufficiently homomorphic.
\subsection{Modular inverses}\label{sec:modular_inversion}
Given a classically known prime $p$ of $n$ bits and an arbitrary superposition of states $\ket{a}$ with $a\in \{1,\dots, p-1\}$. Throughout this section, all arithmetic will be implicitly performed modulo $p$.

First we construct the mask, which is a superposition of integers between $1$ and $p-1$:
\begin{equation}
\frac{1}{\sqrt{p-1}}\ket{a}\sum_{r=1}^{p-1}\ket{r}.
\end{equation}

We then multiply the first and second registers into a third, which masks the value of $a$:
\begin{equation}
\frac{1}{\sqrt{p-1}}\ket{a}\sum_{r=1}^{p-1}\ket{r}\ket{ar}.
\end{equation}
We then measure the third register. Because $p$ is prime, the result will be uniformly random among $\{1,\dots,p-1\}$ and thus reveal no information about $a$. Let $t=ar$. We can then rewrite $r=a^{-1}t$ and our resulting state is $\ket{a}\ket{r}=\ket{a}\ket{a^{-1}t}$.

We then \emph{classically} invert $t$ with the extended Euclidean algorithm. Then we can multiply this value with the second register, to obtain
\begin{equation}
\ket{a}\ket{a^{-1}t}\ket{a^{-1}}
\end{equation}
and finally we multiply the third register by $t$ to clear the second register. 

\paragraph{Costs}
Multiplying by the mask is the most expensive step, since we are multiplying two quantum integers. The best practical circuits for this are $O(n^2)$, which is the same asymptotic cost as the extended Euclidean algorithm~\cite{Proos2003ShorsDL,roetteler_quantum_2017}.

The remaining two multiplications are with classical integers, so with windowing the cost is $O(n^2/\lg n)$~\cite{gidney2019windowed}. 

Overall, we save a constant factor over the extended Euclidean algorithm. This factor could be between 7 and 42, depending on the underlying addition circuits and the cost metric we use~\cite{haner_improved_2020}.

\paragraph{Composite moduli and non-coprime inputs}
If $a=0$ then it has no well-defined inverse. It may be that if we are careful about how we prepare the input to the modular inversion, then it will not have $\ket{0}$ in superposition. However, this may not always be true and we may still wish to ``invert'' an input of $0$ to some specific value (for example, $0$).

For this case, we will instead use a control qubit to just copy the mask $r$ to the third register if $a=0$. Once we have measured the result and inverted it, we also use the check qubit to control the uncomputation of $r$. This allows us to output any value we wish for ``$0^{-1}$'', such as $0$. 

To do the same for thing for non-coprime inputs would require a circuit to detect such inputs; however, this seems to require computing the greatest common divisor, which is as expensive as modular inversion.

If we know the factorization of our modulus, we can construct a superposition of masking integers $r$ that are co-prime to the modulus. When we multiply by our input and measure, then if the result is still co-prime to the modulus, then we have destroyed any states in the superposition that were not coprime to the modulus. Conversely, a non-trivial divisor of the measurement result would imply that we had destroyed any states that were coprime. Which one is preferable will depend on the application.

\paragraph{Uncomputation}
Unlike methods based on the extended Euclidean algorithm, this cannot be made into an in-place algorithm. This is because the measurement means we cannot invert the process. From the final state of $\ket{a}\ket{a^{-1}}$, we would want to multiply $a$ and $a^{-1}$ to clear $a$, but this is unhelpful without \emph{in-place} multiplication: a circuit that maps $\ket{a}\ket{b}$ to $\ket{a}\ket{ab}$. Existing multiplication circuits are out-of-place~\cite{haner_improved_2020,rines2018high,roetteler_quantum_2017}, and a cheap in-place multiplication circuit could simply be inverted to give a cheap division algorithm, making the superposition masking technique unnecessary.

Conversely, a circuit for in-place inversion gives in-place multiplication. Starting from an out-of-place multiplication of $a$ and $b$, which produces registers of $a$, $b$, and $ab$, we would like to clear the register containing $b$. To do this we would invert $a$, then multiply $a^{-1}$ by $ab$ to clear $b$. Then the same inversion circuit can clear $a^{-1}$. Since we cannot clear $a^{-1}$ with our measurement-based technique, the best operation we can create is 
\begin{equation}
\ket{a}\ket{b}\mapsto \ket{a}\ket{a^{-1}}\ket{ab}.
\end{equation}
However, this may be sufficient for certain purposes.

\subsection{Modular Square Roots}
Given an $n$-bit prime $p$, we again start with a superposition of states $\ket{a}$, where all $a$ are assumed to be quadratic residues. We want to compute one of the two integers $b$ such that $b^2\equiv a\mod p$.

We produce the same uniform superposition of masks as for modular inversion. This time, we first square the mask and then multiply with the input, to give
\begin{equation}
\ket{a}\frac{1}{\sqrt{p-1}}\sum_{r=1}^{p-1}\ket{r}\ket{r^2}\ket{ar^2}.
\end{equation}
We then measure $ar^2$. If $a$ is a quadratic residue modulo $p$, then we can find $t:=\pm a^{1/2}r$ from the measurement result. We then uncompute the $\ket{r^2}$ register. The resulting state \ref{eq:sqrt_measurement_result} is still a superposition of $r$, since we could have either $r$ or $-r$:
\begin{equation}\label{eq:sqrt_measurement_result}
\ket{a}\frac{1}{\sqrt{2}}\left(\ket{r}+\ket{-r}\right) = \ket{a}\frac{1}{\sqrt{2}}\left(\ket{a^{-1/2}t}+\ket{-a^{-1/2}t}\right).
\end{equation}
Multiplying the second register by $t^{-1}$ produces 
\begin{equation}
\frac{1}{\sqrt{2}}\ket{a}\left(\ket{a^{-1/2}}+\ket{-a^{-1/2}}\right)
\end{equation}

Depending on the application, we may not want a superposition of the two possible roots. We can remove one of them by comparing each to $\frac{p-1}{2}$ and flipping an ancilla qubit if the state is strictly greater than $\frac{p-1}{2}$. We then use this ancilla to control a modular negation. If we define $a^{-1/2}$ to be the root with value at most $\frac{p-1}{2}$, this process will have the effect of
\begin{equation}
\frac{1}{\sqrt{2}}\left(\ket{a^{-1/2}}+\ket{-a^{-1/2}}\right)\mapsto\ket{a^{-1/2}}\frac{1}{\sqrt{2}}\left(\ket{0}+\ket{1}\right).
\end{equation}
The second register is just a $\ket{+}$ state and can be removed. We are then left with 
\begin{equation}
    \ket{a}\ket{a^{-1/2}}.
\end{equation}

To find $a^{1/2}$, we need to compute a modular inversion. We use the extended Euclidean algorithm. We then square the result to clear the input $\ket{a}$, leaving us with only $\ket{a^{1/2}}$.

\paragraph{Cost}
This costs 2 modular squares, 2 classical-quantum modular multiplications, one modular inversion, and one comparison. Overall, this costs $O(n^2)$ quantum gates. This is better than the $O(n^3)$ gates needed to naively adapt the Tonelli-Shanks algorithm (for worst-case finite fields~\cite{IEEEToC:AdjRod2014}) using the same quantum squaring circuits; however, our total operations still include the classical cost to find the square root.

If we do not need to clear any inputs, we can save one square, and replace the extended Euclidean algorithm with the modular inversion of Section~\ref{sec:modular_inversion}.

\paragraph{Arbitrary roots}
This technique extends to find $a^{1/k}$, given an input of $\ket{a}$, where $k$ is coprime to the group order. The only change is we must produce the state $\ket{r^ka}$ before measuring. This requires $O(\log k)$ modular multiplications, which can be done with limited space using measurement-based pebbling techniques. The total quantum cost would be $O(n^2\log k)$ gates.

\subsection{Sparse Matrix Inversion}
Given a vector $x$ represented as a bitstring in quantum state, we want to compute $A^{-1}x$ for a classical, invertible sparse matrix $A$. 

For a mask, we use a superposition of vectors $r$. We compute the following:
\begin{equation}
    \ket{a}\sum_r \ket{r}\ket{Ar+a}
\end{equation}
If we are in a finite field, we can take $r$ as a superposition over all possible vectors. In other contexts, like vectors in $\R^n$ or $\Z^n$, our mask can be in a superposition of components that are significantly larger than the largest possible value for $a$. This ensures that when we measure $t=Ar+a$, then there is some value of $r$ such that $r=A^{-1}(t-a)$ for all $a$. 

We compute $A^{-1}t=r+A^{-1}a$, and subtract the result from $\ket{r}$ and then negate it. This gives us
\begin{equation}
    \ket{a}\ket{A^{-1}a}.
\end{equation}
We can then uncompute $a$ by multiplying by $A$.

\paragraph{Cost}
If $A$ is an $N\times M$ matrix that is $k$ sparse, this costs $O(Nk)$ quantum multiplications to compute $Ar$ and uncompute $a$, which are the most expensive steps. 

Since $A$ is known classically, we could compute $A^{-1}$ directly and multiply this by $x$. However, a sparse matrix need not have a sparse inverse, so this could cost $O(NM)$ multiplications instead. Thus, the technique saves $O(N(M-k))$ quantum multiplications.

We caution that this technique applies only to vectors represented as bitstrings, and is thus unrelated and inapplicable to quantum linear algebra techniques based on superpositions such as~\cite{HHL2009}.
\subsection{Group Homomorphisms}
As one possible generalization, let $G$ be a group for which there are quantum circuits to produce a uniform superposition of elements in $G$, and to perform the group law. Let $f$ be a homomorphic, invertible function on $G$ for which we also have a quantum circuit. We start with a state $\ket{a}$ and we want to find $f^{-1}(a)$. We will describe a procedure to produce $(f^{-1}(a))^{-1}$.

Starting with state $\ket{a}$, the masks are a uniform superposition of elements of $G$:
\begin{equation}
\ket{a}\frac{1}{\sqrt{\vert G\vert}}\sum_{r\in G}\ket{r}
\end{equation}
We then compute $f(r)$ in another register, then multiply this by $a$:
\begin{equation}
\frac{1}{\sqrt{\vert G\vert}} \ket{a}\sum_{r\in G}\ket{r}\ket{f(r)}\ket{af(r)}.
\end{equation}
Then we can uncompute $f(r)$ and measure $af(r)$. Let $x=f^{-1}(a)$. Since $f$ is homomorphic, we have $af(r)=f(xr)$. Since we know this value, we classically invert $f$ to find $xr$. If we denote $t=xr$, we can rewrite our quantum state as 
\begin{equation}
\ket{a}\ket{x^{-1}t}
\end{equation}
Then we use the group operation with $t^{-1}$ and then $t$, as in the previous examples, to get $\ket{a}\ket{x^{-1}}$.

The total cost is one evaluation of the function $f$, one quantum-quantum group operation, two classical-quantum group operations, and the classical cost to invert $f$. 

In the modular inversion case, $f$ was the identity; the inverse is a side-effect of the general technique. For square roots, $f(x)=x^2$, and for general roots $f(x)=x^n$. Matrix inversion uses $f(x)=Ax$, where the additive inverse is easy to compute.

This generalization shows that our technique could help with homomorphic functions on real numbers, such as roots, logarithms, and even inverse trigonometric functions.

For example, to compute $x$ from $y=\sin(x)$, we can compute $\cos(x)=\sqrt{1-y^2}$, create a mask $r$ then compute and measure 
\begin{equation}
\sin(x)\cos(r)+\cos(x)\sin(r)=\sin(x+r).
\end{equation}

We can then invert this classically to get $x+r$.

\subsection{In-place Division}
Given a register of states $\ket{a}$ of $n$ bits, and a classical argument $b$, our goal is to produce $\ket{\floor{a/b}}\ket{a\mod b}$.

The mask is a superposition of $r_1$ and $r_2$, where $r_1$ ranges from $0$ to $2^m-1$ and $r_2$ ranges from $0$ to $b-1$. We will parameterize $m$ at the end. Define $r:=r_1b+r_2$; we add $r$ to $a$ to get
\begin{equation}
    \ket{a+r}\ket{r_1}\ket{r_2}
\end{equation}
and then we measure $a+r$.

We can represent $a$ uniquely as $a=a_1b+a_2$, where $a_1=\floor{a/b}$ and $a_2=a\mod b$. Similarly, $a+r=c_1b+c_2$. We have the following facts:
\begin{align}
    c_1 = a_1+r_1 &\Leftrightarrow a_2+r_2 < b\label{eq:modular_condition_1}\\
    c_1 = a_1 + r_1+1 &\Leftrightarrow a_2+r_2 \geq b \label{eq:modular_condition_2}
\end{align}
We also know that $c_2\equiv a_2+r_2\mod b$. and so $c_2-r_2=a_2$ if and only if $c_2-r_2\geq 0$; otherwise, $c_2-r_2+b=a_2$. These are the same conditions as \ref{eq:modular_condition_1} and \ref{eq:modular_condition_2}.

Since we can compute $c_1$ and $c_2$ classically, we compute $c_2-r_2$ in the register for $r_2$ and $c_2-r_1$ in the register for $r_1$. We check if $c_2-r_2$ is negative; if it is, we add $b$ to that register and add $1$ to $c_r-r_1$. This gives us $\ket{a_1}\ket{a_2}$. To clear the comparison qubit, we check if $c_2-a_2$ is negative.

This carries some probability of failure, since a basic $N$-bit modular adder will add modulo $2^N$. We need $a+r\leq 2^N$, which means that if we measure $c=a+r$, then for every value of $a\leq c$ in the superposition, there is precisely 1 value of $r$ such that this holds. If $a>c$, there are $0$ such values of $r$. Thus, as long as our measured result is not smaller than the largest value of $a$ in superposition, we do not change the state at all. Since $a < 2^n$ and $r$ is approximately uniformly random among $m+\floor{\lg b}$-bit integers, the probability of failure is $2^{n-m-\floor{\lg b}}$.

\paragraph{Cost}
The cost here is dominated by the multiplication $r_1\times b$, which we add directly into the register with $a$. We need $m$ additions of a classical integer into an $N:=m+\floor{\lg b}$-bit register, and since $b$ is classical these additions can be windowed~\cite{gidney2019windowed}. This means the total cost is $O(mN/\lg N)$.

Rines and Chuang~\cite{rines2018high} provide a circuit for the same task that requires $2(n-\floor{\lg b})$ additions, where $n$ is the length of the initial quantum register. The additions range from $n$ bits to $\floor{\lg b}$ bits, for a total cost of $O(n^2)$. Since we only need to take $m$ as a constant multiple of $n$ for for exponentially suppressed error, our technique is asymptotically cheaper.
\section{Discussion}\label{sec:discussion}
\paragraph{Applicable functions:}
For any function that we use with this technique, we must have a classical method to compute it. With generic transformations, we could transform this into a quantum circuit with only a constant overhead in gates. Thus this technique is only applicable in a specific context, where we have non-asymptotic cost goals. The constant factor difference between classical and quantum computation puts us in a strange place, where asymptotics do not reflect the best implementations.

In all of our applications, to compute a function $f$ on the input we needed to apply $f^{-1}$ to the mask. Hence, our technique only helps with functions that are at least partially one-way, and the improvement is greater for functions that have a large gap in efficiency between $f$ and $f^{-1}$. However, every strongly one-way function that we can think of is either insufficiently homomorphic (e.g., cryptographic hash functions) or the inverse function is much easier for the quantum computer to compute (e.g., group exponentiation). 

\paragraph{Uncomputation:}
The greatest problem with this technique is that by introducing measurements, we have moved out of a pure quantum circuit model, so we cannot invert this process. This means that typical techniques like Bennett's reduction do not apply. In particular, the technique is inherently out-of-place. For modular inversion, most applications call for a circuit to compute
\begin{equation}
    \ket{a}\mapsto\ket{a^{-1}}
\end{equation}
but our method computes
\begin{equation}
    \ket{a}\mapsto\ket{a}\ket{a^{-1}}.
\end{equation}

With a quantum circuit, we would apply the inverse circuit with the roles of $a$ and $a^{-1}$ switched, which would uncompute $a$. With our method, there is no inverse circuit. The only way to uncompute $a$ from $a^{-1}$ is to use an expensive quantum circuit like the extended Euclidean algorithm, but this defeats the cost savings of the masking technique. For this reason, we are unable to use our technique to provide any improvement to elliptic curve point addition, which requires uncomputing modular inverses.

\paragraph{Complexity}
Jozsa conjectured that interleaving polylogarithmic-depth quantum computation with a polynomial-size classical computation can simulate any polynomial-time quantum computation~\cite{jozsa2006}. While recent work provided an oracle separation between these classes~\cite{chia2019need,coudron2019computations}, superposition masking provides a specific tool that may be able to simulate higher-depth quantum algorithms with a high-depth classical oracle.


\begin{thebibliography}{10}

\bibitem{IEEEToC:AdjRod2014}
G.~{Adj} and F.~{Rodríguez-Henríquez}.
\newblock Square root computation over even extension fields.
\newblock {\em IEEE Transactions on Computers}, 63(11):2829--2841, 2014.

\bibitem{Berry2019qubitizationof}
Dominic~W. Berry, Craig Gidney, Mario Motta, Jarrod~R. McClean, and Ryan
  Babbush.
\newblock Qubitization of {A}rbitrary {B}asis {Q}uantum {C}hemistry
  {L}everaging {S}parsity and {L}ow {R}ank {F}actorization.
\newblock {\em {Quantum}}, 3:208, December 2019.

\bibitem{bravyi_kitaev_2004}
Sergey Bravyi and Alexei Kitaev.
\newblock Universal quantum computation with ideal clifford gates and noisy
  ancillas.
\newblock {\em Phys. Rev. A}, 71:022316, Feb 2005.

\bibitem{SciAdv:Brown2020}
Benjamin~J. Brown.
\newblock A fault-tolerant non-clifford gate for the surface code in two
  dimensions.
\newblock {\em Science Advances}, 6(21), 2020,
  https://advances.sciencemag.org/content/6/21/eaay4929.full.pdf.

\bibitem{chia2019need}
Nai-Hui Chia, Kai-Min Chung, and Ching-Yi Lai.
\newblock On the need for large quantum depth, 2019
arxiv:1909.10303


\bibitem{coudron2019computations}
Matthew Coudron and Sanketh Menda.
\newblock Computations with greater quantum depth are strictly more powerful
  (relative to an oracle), 2019
  arxiv:1909.10503

\bibitem{FMMC2012surface_code}
Austin~G. Fowler, Matteo Mariantoni, John~M. Martinis, and Andrew~N. Cleland.
\newblock Surface codes: Towards practical large-scale quantum computation.
\newblock {\em Phys. Rev. A}, 86:032324, Sep 2012.

\bibitem{FWH2012surface_code}
Austin~G. Fowler, Adam Whiteside, and Lloyd Hollenberg.
\newblock Towards practical classical processing for the surface code.
\newblock {\em Physical review letters}, 108:180501, 05 2012.

\bibitem{gheorghiu2019benchmarking}
Vlad Gheorghiu and Michele Mosca.
\newblock Benchmarking the quantum cryptanalysis of symmetric, public-key and
  hash-based cryptographic schemes, 2019
  arxiv:1902.02332

\bibitem{gidney2019windowed}
Craig Gidney.
\newblock Windowed quantum arithmetic, 2019, 
arxiv:1905.07682

\bibitem{gidney2019factor}
Craig Gidney and Martin Ekerå.
\newblock How to factor 2048 bit {RSA} integers in 8 hours using 20 million
  noisy qubits, 2019, 
  arxiv:1905.09749

\bibitem{grover_1996}
Lov~K. Grover.
\newblock A fast quantum mechanical algorithm for database search.
\newblock In {\em Proceedings of the Twenty-Eighth Annual ACM Symposium on
  Theory of Computing}, STOC ’96, page 212–219, New York, NY, USA, 1996.
  Association for Computing Machinery.

\bibitem{HHL2009}
Aram~W. Harrow, Avinatan Hassidim, and Seth Lloyd.
\newblock Quantum algorithm for linear systems of equations.
\newblock {\em Phys. Rev. Lett.}, 103:150502, Oct 2009.

\bibitem{horsman_surface_2012}
Clare Horsman, Austin~G. Fowler, Simon Devitt, and Rodney~Van Meter.
\newblock Surface code quantum computing by lattice surgery.
\newblock {\em New Journal of Physics}, 14(12):123011, December 2012.

\bibitem{haner_improved_2020}
Thomas Häner, Samuel Jaques, Michael Naehrig, Martin Roetteler, and Mathias
  Soeken.
\newblock Improved {Quantum} {Circuits} for {Elliptic} {Curve} {Discrete}
  {Logarithms}.
\newblock In Jintai Ding and Jean-Pierre Tillich, editors, {\em Post-{Quantum}
  {Cryptography}}, pages 425--444, Cham, 2020. Springer International
  Publishing.

\bibitem{jones_2013}
Cody Jones.
\newblock Low-overhead constructions for the fault-tolerant toffoli gate.
\newblock {\em Phys. Rev. A}, 87:022328, Feb 2013.

\bibitem{jozsa2006}
Richard Jozsa.
\newblock An introduction to measurement based quantum computation.
\newblock In Dimitris~G. Angelakis, Matthias Christandl, Artur Ekert, Alastair
  Kay, and Sergei Kulik, editors, {\em Quantum Information Processing - From
  Theory to Experiment}, NATO Science Series {III}: Computer and Systems
  Sciences., chapter~2, pages 137--158. 2006.

\bibitem{Proos2003ShorsDL}
John Proos and Christof Zalka.
\newblock Shor's discrete logarithm quantum algorithm for elliptic curves.
\newblock {\em Quantum Information \& Computation}, 3:317--344, 2003.

\bibitem{rines2018high}
Rich Rines and Isaac Chuang.
\newblock High performance quantum modular multipliers, 2018
arxiv:1801.01081

\bibitem{roetteler_quantum_2017}
Martin Roetteler, Michael Naehrig, Krysta~M. Svore, and Kristin Lauter.
\newblock Quantum {Resource} {Estimates} for {Computing} {Elliptic} {Curve}
  {Discrete} {Logarithms}.
\newblock In Tsuyoshi Takagi and Thomas Peyrin, editors, {\em Advances in
  {Cryptology} – {ASIACRYPT} 2017}, pages 241--270, Cham, 2017. Springer
  International Publishing.

\bibitem{shor_1994}
Peter~W. Shor.
\newblock Polynomial time algorithms for discrete logarithms and factoring on a
  quantum computer.
\newblock In Leonard~M. Adleman and Ming-Deh Huang, editors, {\em Algorithmic
  Number Theory}, pages 289--289, Berlin, Heidelberg, 1994. Springer Berlin
  Heidelberg.

\bibitem{suchara_et_al_2013}
M.~{Suchara}, J.~{Kubiatowicz}, A.~{Faruque}, F.~T. {Chong}, C.~{Lai}, and
  G.~{Paz}.
\newblock Qure: The quantum resource estimator toolbox.
\newblock In {\em 2013 IEEE 31st International Conference on Computer Design
  (ICCD)}, pages 419--426, 2013.

\bibitem{TMNCQM2017QECC_hardware}
Swamit~S. Tannu, Zachary~A. Myers, Prashant~J. Nair, Douglas~M. Carmean, and
  Moinuddin~K. Qureshi.
\newblock Taming the instruction bandwidth of quantum computers via
  hardware-managed error correction.
\newblock In {\em Proceedings of the 50th Annual IEEE/ACM International
  Symposium on Microarchitecture}, MICRO-50 ’17, page 679–691, New York,
  NY, USA, 2017. Association for Computing Machinery.

\bibitem{vanmeter2005exp}
Rodney Van~Meter and Kohei~M. Itoh.
\newblock Fast quantum modular exponentiation.
\newblock {\em Phys. Rev. A}, 71:052320, May 2005.

\end{thebibliography}
\end{document}